# A Dynamic Theory for Explaining the Evaporation Paradox and Global Energy Transpiration

Kejing Liu[a,b*]

[a] *State Key Laboratory of Hydroscience and Engineering, Tsinghua University, Beijing, China*

[b]*Jimei University, Fujian, China*

*Corresponding author address:* Kejing Liu, liu_kj313@126.com

ABSTRACT

Evaporation is a part of water cycle and a process of energy exchange between atmosphere and land surface, its variation reflects global change. Pan evaporation decreases with global warming is the phenomena named evaporation paradox, which exits in worldwide and spatio – temporal. Broad-scale observations between the 1950s and 2000s revealed that global pan evaporation ($E_{pan}$) decreases with increasing quantity of clouds. However, in the Huaihe River Basin, both the total cloud quantity and $E_{pan}$ decreased during this period, and similar phenomena were observed in some other regions of the globe. A nonlinear second-order neutral-delay dynamic equation (NSNDE) of the change in the cloud quantity and $E_{pan}$ with time was constructed, encompassing different stages (formation, duration, waning, recurring) of the evaporation paradox. On the basis of this equation, a new model named 'steamer' was proposed, encompassing a set of dynamic equations to explore the evaporation paradox. The effects of the total cloud quantity on factors that affect the sensible heat flux are investigated, revealing that actual evaporation ($E_a$) displays similar oscillation properties as $E_{pan}$ and the total cloud quantity, and their relationship is complimentary in some stages of the evaporation paradox. On the basis of the relation between the total cloud quantity and evaporation, an expression for global energy transpiration was established, and the time delay plays an important role in energy exchange between global spheres. This relation indicates the stability of atmosphere and surface.

SIGNIFICANCE STATEMENTS

This study provides a model based on a nonlinear second-order neutral-delay dynamic equation to explore the evaporation paradox (phenomenon of evaporation decreasing as temperature increases) and global energy transpiration, which is a multistage process. The findings indicate that the cloud quantity and evaporation form a dynamic system that is spatiotemporally variable, and this relationship within the land–atmosphere system is unstable.

## 1. Introduction

In the past 50 years, a steady decline in pan evaporation ($E_{pan}$) has been observed worldwide, while the temperature continues to increase. This phenomenon is called the 'evaporation paradox'. Research before 2010 revealed that an obvious evaporation paradox existed in northern European Russia (taiga) (1951—1990), southern European China (1951—1987), Siberia (taiga) (1951—1988) (Golubev et al. 2001), the southern and southwestern USA (1940—2000) (Lawrimore and Peterson 2000), western Northwest China (1957—2001) (Qiu et al. 2003), the lower reaches of the Yangtze River basin (1960—2000) (Wang et al. 2005), southern China (1956—2005) (Cong et al. 2009), and Australia (1975—2005) (Roderick and Farquhar 2004). An obvious evaporation paradox existed widely from the 1960s—1980s. However, $E_{pan}$ has not decreased everywhere or continuously in time; it has increased in Florida in the USA (1940—2000) (Lawrimore and Peterson 2000), northeastern China (1956—2005) (Cong et al. 2009), and the middle reaches of the Yangtze River basin (1960—2000) (Wang et al. 2005).

There are two main explanations for the evaporation paradox. One is based on the hydrological cycle, suggesting that a decrease in $E_{pan}$ represents an increase in actual evaporation ($E_a$), which is called the complementary relationship (Brutsaer and Parlanger 1998). Another more popular explanation is that the decline in $E_{pan}$ is caused mainly by global solar irradiance decreases resulting from increased cloud coverage; their relationship can be described as a negative correlation (Roderick and Farquhar 2002). This theory has been proven by universal observations, but not all records worldwide support it. Evaporation is an important part of energy exchange among global spheres, and the divergences among explanations of evaporation reveal that theories about global energy and heat transportation are incomplete (Lawrimore and Peterson 2000; Golubev et al. 2001; Qiu et al. 2003; Roderick and Farquhar 2004; Wang et al. 2005; Cong et al. 2009).

Surface station-based observations reveal that the mean total cloud quantity increased before the mid-1980s or early 1990s. However, International Satellite Cloud Climatology Project (ISCCP) data show that it decreased after the early 1990s. Fig. 1 (Fig. 1(a)) shows the surface observations of total cloud cover (ocean only) between 1952 and 1997 from the Extended Edited Cloud Report Archive (EECRA). Fig. 1(b) shows the total cloud quantity between 1983 and 2004 from the ISCCP, with oscillations in the total cloud quantity between the 1950s and 2000s. Fig. 2 represents the spatial distribution trends of low clouds (a), middle clouds (b) and high clouds (c) from 1983—2001 of the ISCCP D2 dataset (Ding et al. 2004); positive values represent an increasing trend of the clouds, and negative values represent a decreasing trend. In total, low clouds and high clouds decreased in abundance, whereas the quantity of middle clouds increased slightly during this period. Closer observations revealed that the total cloud quantity increased in northern Russia, European Russia, Siberia (taiga), the northwestern USA, most parts of Australia and the lower reaches of the Yangtze River basin in China, whereas it decreased in eastern Inner Mongolia, Northeast China and the Jiaodong Peninsula in Shandong, China, southern Europe, the southeastern USA, most parts of Mexico, eastern Brazil, the Arabian Peninsula, eastern Africa and Madagascar (Ding et al. 2004).

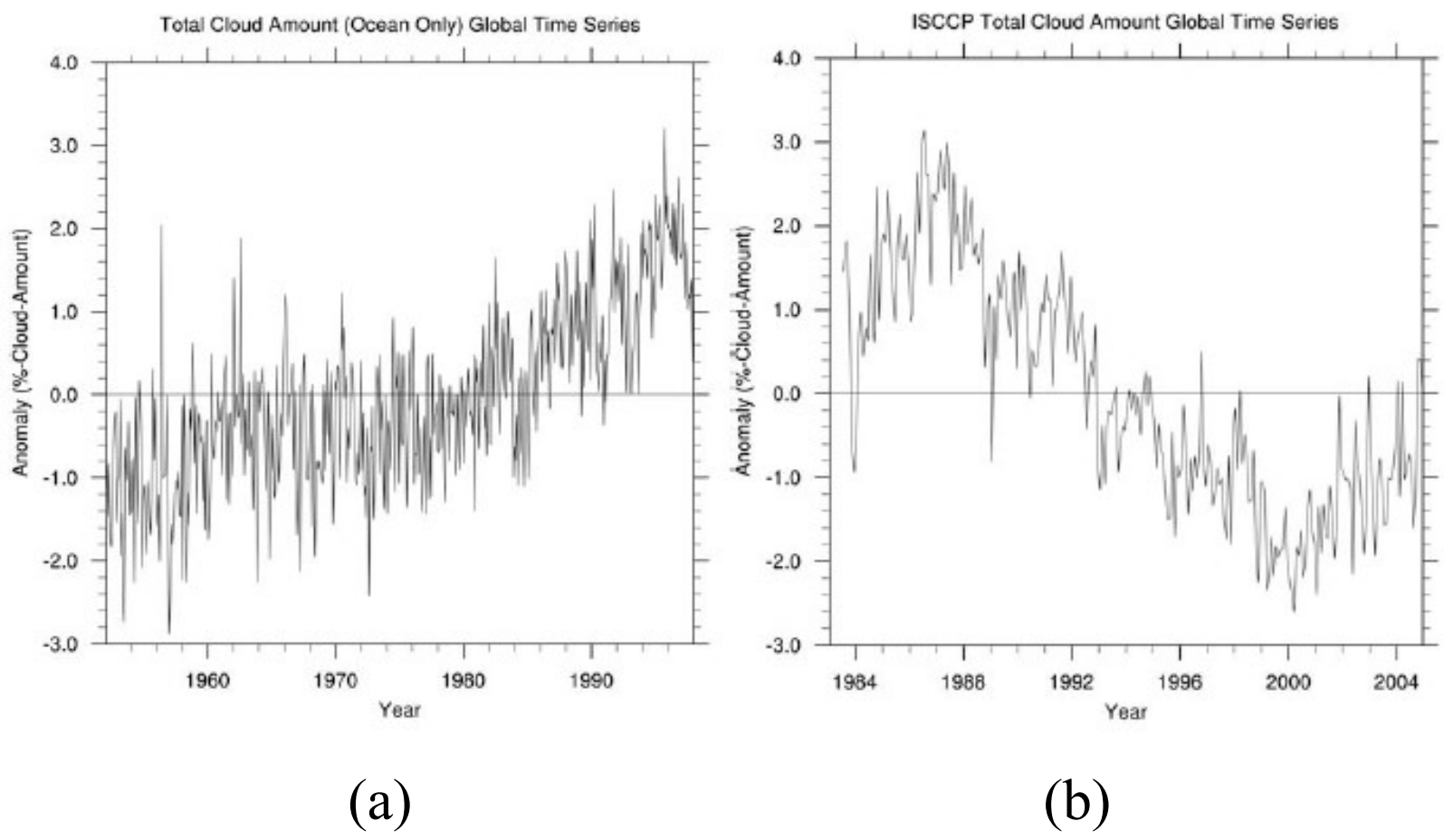

(a) (b)

Fig. 1. (a) 1952−1997 surface observations of total cloud cover from the Extended Edited Cloud Report Archive (EECRA), created by S. Warren and C. Hahn. (b) 1983−2004 total cloud quantity from the ISCCP, created by the W. Rossow group

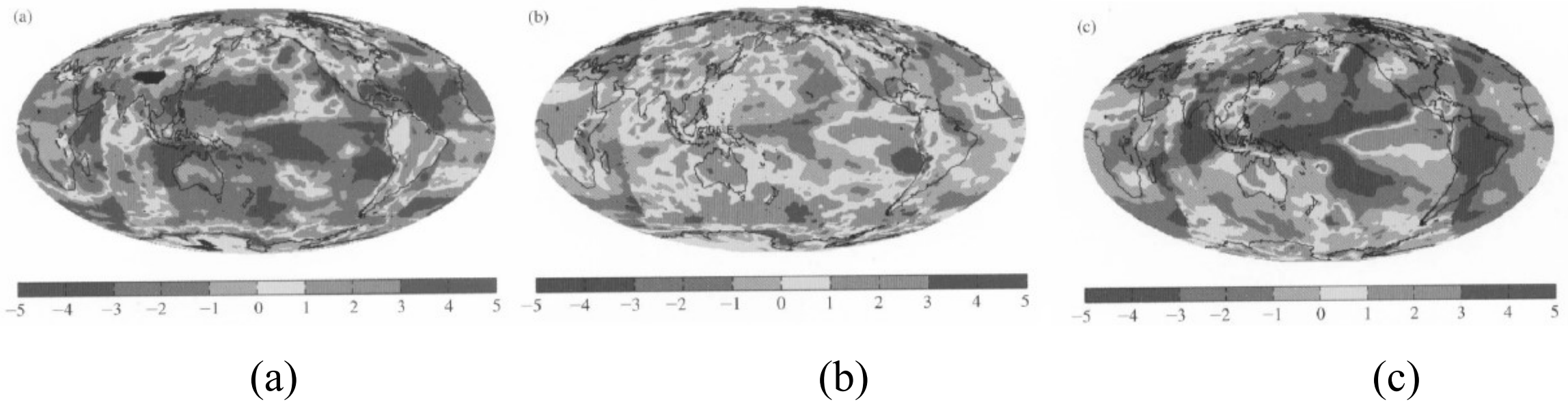

(a) (b) (c)

Fig. 2. 1983−2001 cloud quantity variation distributions; created by Ding et al. (2004). The positive values indicate where cloud quantity increase and the negative values indicate where it decrease, the unit is %/10 year.

On the basis of the temporal and spatial distributions of the cloud quantity and the $E_{pan}$ trend distributions, in the observed period, in most parts of Europe, the USA and Australia, $E_{pan}$ and cloud quantity increased, whereas they decreased in most parts of China. Thus, the relationship between cloud quantity and $E_{pan}$ cannot be described as a simple negative correlation. Moreover, evaluating actual evaporation is very difficult and costly, and most relations are theorized on an idealized moist environment; another explanation of the evaporation paradox, the complementary relationship, is not suitable for every situation worldwide.

Many researchers have investigated the factors affecting the decrease in $E_{pan}$ worldwide, often focusing on the effects of increasing relative humidity. N. Chattopadhyay and M. Hulme (1997) analyzed historical evaporation data and reported that relative humidity has a strong relationship with changes in $E_{pan}$ and potential evapotranspiration (*PE*). M. T. Hobbins et al. (2004) reported that in the conterminous U.S. from 1950—2002, a decreasing evaporation trend was observed at 64% of the 44 annual pans deployed, and $E_{pan}$ had a negative correlation with $E_a$. They reported that incident solar radiation provides the major energy input for any evaporative process and that humidity is a key component of $E_a$. D.H. Burn and N. M. Hesch (2007) indicated a decrease in the wind speed or an increase in the vapor pressure deficit can be treated as an increase in atmospheric moisture, which causes a decrease in $E_{pan}$. Other studies have shown that a decrease in solar irradiance is the driving force of a decrease in $E_{pan}$ (Liu et al. 2004). In some regions, such as China, a decrease in solar irradiance is not always accompanied by an increase in cloud quantity, but aerosols may play a critical role in the $E_{pan}$ trend. In addition to radiation, a decrease in the temperature daily range (TDR) is another important factor, e.g., a decrease in $E_{pan}$ in China was found to decrease solar irradiance given the influence of aerosols (Ren et al. 2006; Qian et al. 2006).

Changes in meteorological factors are strongly linked with changes in cloud conditions (Albert Arking, 1991). The mean effect of clouds on the Earth is cooling, although the magnitude varies, and estimates of temperature sensitivity to the quantity of clouds vary worldwide. The sensitivity to cloud condensation nuclei (CCNs) shows that aerosols associated with $SO_2$ emissions can lead to a reduction in solar heating. Peterson et al. (1995) noted the direct link between decreases in $E_{pan}$ and cloud coverage, which is an implication of decreasing radiation. They also noted that increases in cloud cover may account for decreases in TDR and that TDR is highly correlated with decreases in $E_{pan}$. Thus, cloud cover has both direct and indirect effects on Epan changes. Young and Sabburg (2006) reported that clouds are negatively correlated with solar radiation and temperature and positively correlated with humidity at the Earth's surface. Research has shown that cloud coverage reduces temperature and increases relative humidity (RH). A decreasing temperature and increasing RH result in reduced evapotranspiration estimates (*ET*) via the Penman–Monteith equation. Farquhar and Roderick (2003) reported that increasing aerosol emissions from pollution caused the sky to become cloudier and inhibited evaporation from the land surface.

The present research shows that key meteorological forces affecting the decrease in $E_{pan}$ vary globally, and the relationships among these factors are poorly understood. Moreover, there are different explanations for the universal $E_{pan}$ phenomenon, and each is suitable in certain situations, suggesting that is a complex phenomenon with different patterns and processes. Although many studies have revealed that cloud cover affects meteorological factors, further studies are needed to determine the link between $E_{pan}$ changes and cloud trends and to resolve the conflicts between the present theories.

This study is based on observations of $E_{pan}$ and cloud quantity changes in the Huaihe River Basin of China for the region between the Yangtze River basin and the Yellow River Basin, which is a transitional zone of South and North of China in climate and with high complexity, and give the relationships between evaporation and cloud quantity variation at different scales and stages by a set of equations based on dynamic system are explored to provide a complete explanation of the evaporation paradox, including the relationships among $E_{pan}$, $E_a$ and cloud quantity. The relations reflected energy transpiration between atmosphere and land surface.

## 2. Data

The Huaihe River Basin, situated in eastern China between the Yangtze and Yellow River valleys, spans approximately 270,000 km² across Henan, Anhui, Jiangsu, and Shandong provinces. This transitional zone features a unique warm temperate monsoon climate with distinct seasonal variations. Annual precipitation averages 800-1,000 mm, concentrated in summer months (June-September), while winters remain relatively dry. The basin's flat alluvial plains (elevation below 50m) contrast with western hilly areas reaching 200-500m, creating diverse microclimates. Dominated by fertile loess soils and crisscrossed by dense tributary networks, the region supports intensive agriculture, particularly wheat and rice cultivation. However, its low-lying topography (much land below river levels) and seasonal rainfall concentration make it historically prone to devastating floods. The basin's 1,100-km main stem flows eastward from the Tongbai Mountains to Hongze Lake, ultimately draining into the Yellow Sea via Yangtze River connections. Notable geographical features include numerous artificial reservoirs, flood diversion projects, and the north-south transitional vegetation blending deciduous broadleaf and evergreen forests.

The surface-based datasets for cloud quantity (denoted as $Q$) and $E_{pan}$ in the Huaihe River Basin used in this study were obtained from the China Meteorological Administration (http://www.cdc.cma.gov.cn/), In the records, $Q$ is presented as the annual average total cloud quantity (0.1 Okta/year), $E_{pan}$ is presented as the annual total pan evaporation (mm/year), and the records span the period of 1954—2005. Notably, the data were recorded at 14 national standard meteorological stations: Yanzhou (54916), Juxian (54936), Xuchang (57089), Kaifeng (57091), Dangshan (58015), Xihua (57193), Xinyang (57297), Bozhou (58102), Suzhou (58122), Sheyang (58150), Fuyang (58203), Bengbu (58221), Dongtai (58251) and Huoshan (58314). The research area and the stations are show in the Fig. 3.

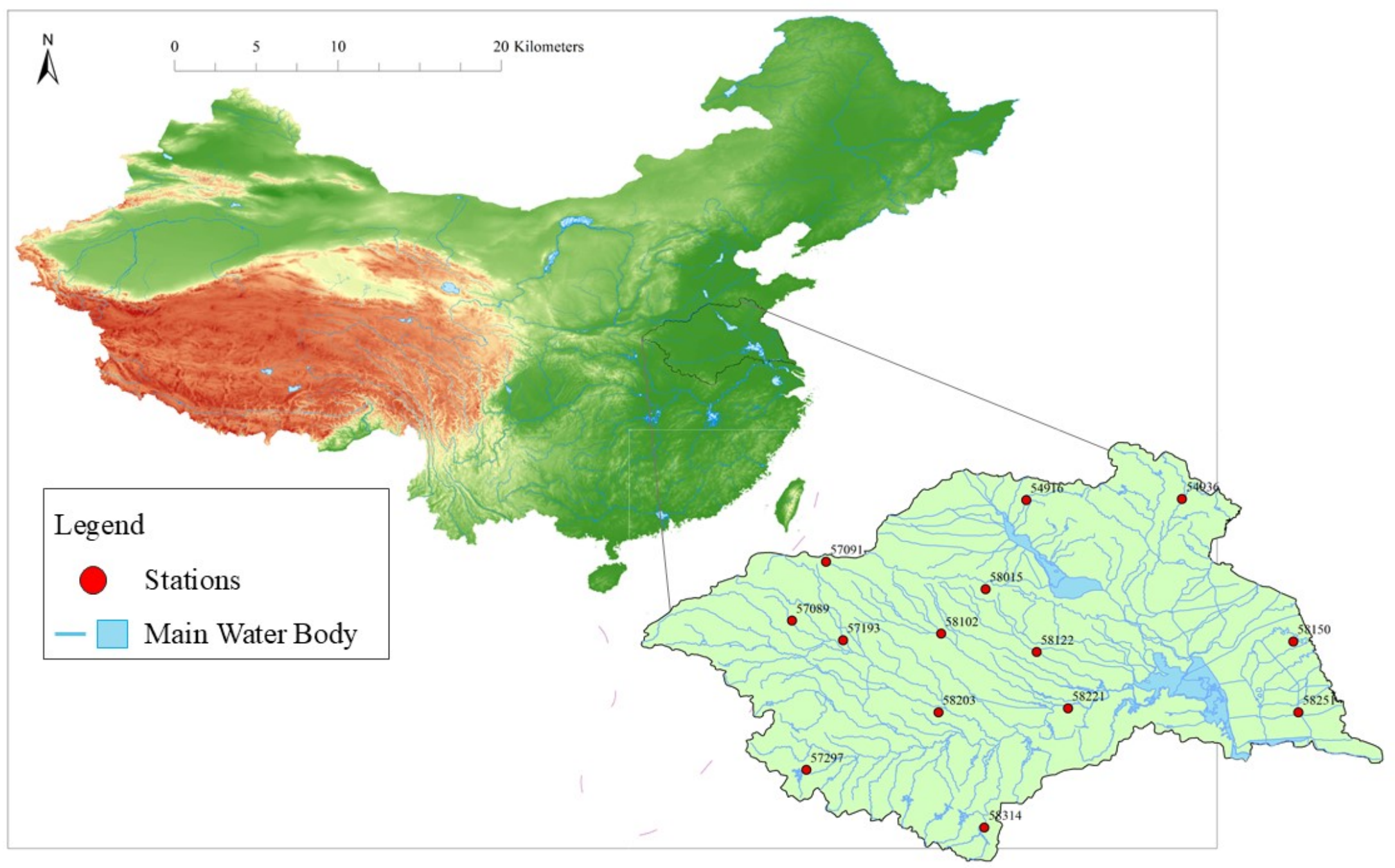

Fig.3 Huaihe River Basin and the national standard meteorological stations

## 3. Methods and Analysis

*a). Trends and abrupt changes in cloud quantity and evaporation*

The Mann–Kendall (M-K) test, a nonparametric trend detection method widely used in hydrology (Kendall 1975), was applied to the data from 1954—2005, and the results revealed that $Q$ and $E_{pan}$ presented similar declining trends at most stations in the Huaihe River Basin. During the same period, the temperature clearly increased in the area, and an evaporation paradox occurred. To determine the maximum step change points of $Q$ and $E_{pan}$, Bernaola-Galván (B-G) test (Bernaola-Galván et al. 2001) were employed; specifically, a heuristic B-G segmentation algorithm was used for change point (where the maximum mean change occurs) detection.

In addition to identifying abrupt changes in the whole series, $Q$ and $E_{pan}$ may present different correlations at different time scales, and the different properties that exist at various scales may aid in deriving the general characteristics of the relationships between evaporation and clouds. Thus, the annual ranges of changes, $dx/dt$, in $Q$ and $E_{pan}$ were considered.

1) MANN-KENDALL (M-K) TEST

The M-K test is based on the statistic $S$ defined by:

$$S = \sum_{i=1}^{n-1} \sum_{j=i+1}^{n} \mathrm{sgn}\left(x_j - x_i\right) \tag{1}$$

where $x_i$ and $x_j$ are the sequential data values, $n$ is the length of the series, and

$$\mathrm{sgn}\left(\theta\right) = \begin{cases} 1 & ,\text{if } \theta > 0 \\ 0, & \text{if } \theta = 0 \\ -1, & \text{if } \theta < 0 \end{cases} \tag{2}$$

The statistic $S$ is approximately normally distributed with the mean and the variance as follows:

$$E\left(S\right) = 0 \tag{3}$$

$$V\left(S\right) = \frac{n\left(n-1\right)\left(2n+5\right) - \sum_{i=1}^{n} t_i i\left(i-1\right)\left(2i+5\right)}{18} \tag{4}$$

where $t_i$ is the number of ties of extent $i$. The standardized test statistic $Z$ is computed by

$$Z_{MK} = \begin{cases} \dfrac{S-1}{\sqrt{\mathrm{Var}\left(S\right)}}, & S > 0 \\ 0, & S = 0 \\ \dfrac{S+1}{\sqrt{\mathrm{Var}\left(S\right)}}, & S < 0 \end{cases} \tag{5}$$

The standardized M-K statistic $Z_{MK}$ follows the standard normal distribution with mean zero and variance of one.

The $P$-value (probability value) of the $S$ can be estimated using the normal CDF,

$$p = 0.5 - \Phi\left(\left|Z_{MK}\right|\right) \tag{6}$$

where

$$\Phi\left(\left|Z_{MK}\right|\right) = \frac{1}{\sqrt{2\pi}} \int_0^{\left|Z_{MK}\right|} e^{-t^2/2} dt$$

If the $P$-value is small enough, the trend is quite unlikely to be caused by random sampling. For example, at significance level of 0.05, if $p \le 0.05$, the existing trend is considered to be statistically significant.

2) BERNAOLA-GALVÁN (B-G) TEST

The statistic $t$ is computed by

$$t_i = \left| \frac{\mu_i^{left} - \mu_i^{right}}{S_D} \right| \quad (7)$$

where $i$ is every position of a sliding point from left to right along the signal, $\mu_i^{left}$ is the mean of the subset of the signal to the left of the point and $\mu_i^{right}$ is to the right, and

$$S_D = \sqrt{\left( \frac{1}{N_{left}} + \frac{1}{N_{right}} \right) \left( \frac{s_{left}^2 + s_{right}^2}{N_{left} + N_{right} - 2} \right)} \quad (8)$$

is the pooled variance, $s_{left}$ and $s_{right}$ are the standard deviations of the data to the left and to the right of the point, $N_{left}$ and $N_{right}$ are the number of points to the left and to the right of the sliding point.

Point $p$ with maximum value of $t$ is checked by a modified $T$-test, its significance level $P$ is defined by

$$P(t_{\max}) = \text{Prob}\{t \le t_{\max}\} \quad (9)$$

For which cannot be obtained in a closed analytical form, so an approximation by Monte Carlo simulations

$$P(t_{\max}) \approx \left\{ 1 - I_{\left[v/\left(v+t_{\max}^2\right)\right]}(\delta v, \quad \delta) \right\}^{\gamma} \quad (10),$$

where $\gamma = 4.19 \ln N - 11.54$, $\delta = 0.40$, $N$ is the length of the series, $v = N - 2$ is the number of degrees of freedom, and $I_x(a, \quad b)$ is the incomplete beta function.

If $P$ exceeds a selected threshold $P_0$, the series is cut into two subsequences at this point, and this point is an abrupt change point; otherwise, the series remains undivided. If the sequence is cut, the procedure continues recursively for each of the two resulting subsequences created by each cut. The process stops when none of the possible cutting points has significance exceeding $P_0$.

The $Q$ series, $E_{pan}$ series, annual ranges of changes and change detection results are shown in Fig. 4.

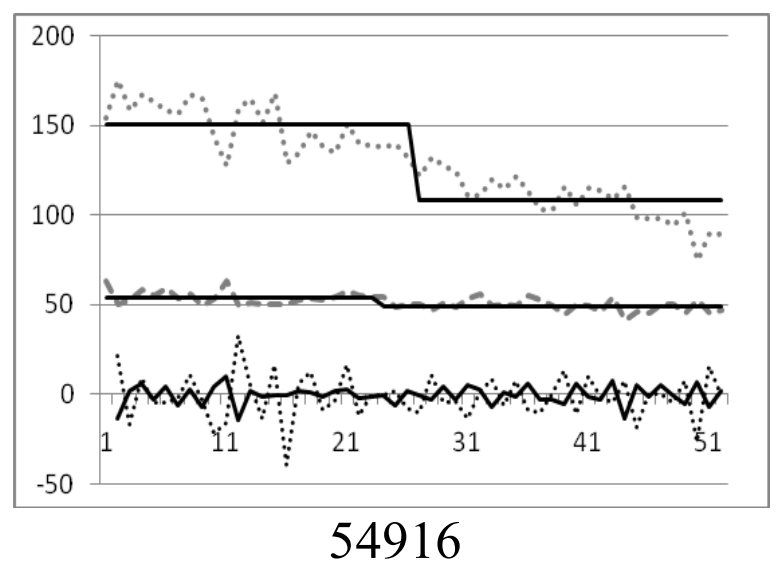

54916

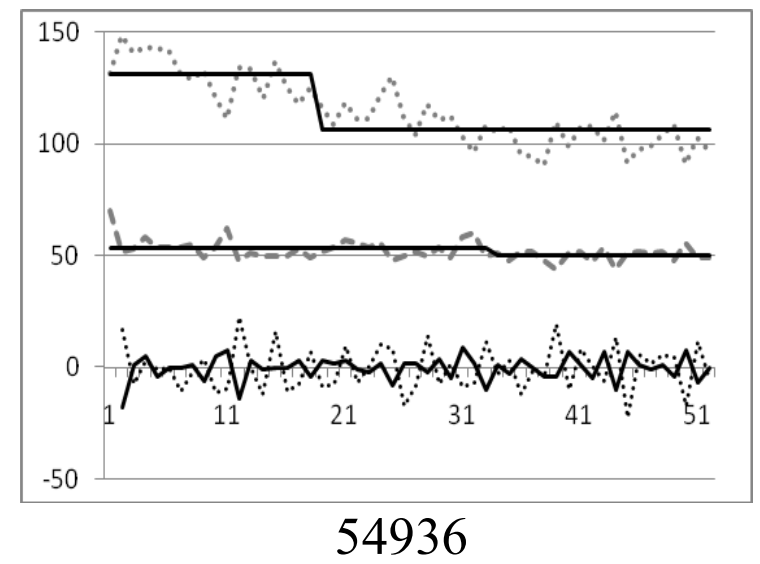

54936

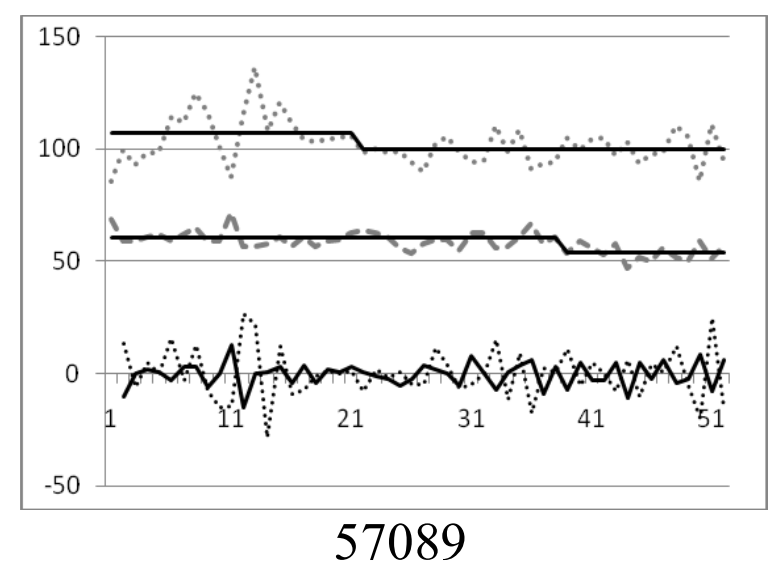

57089

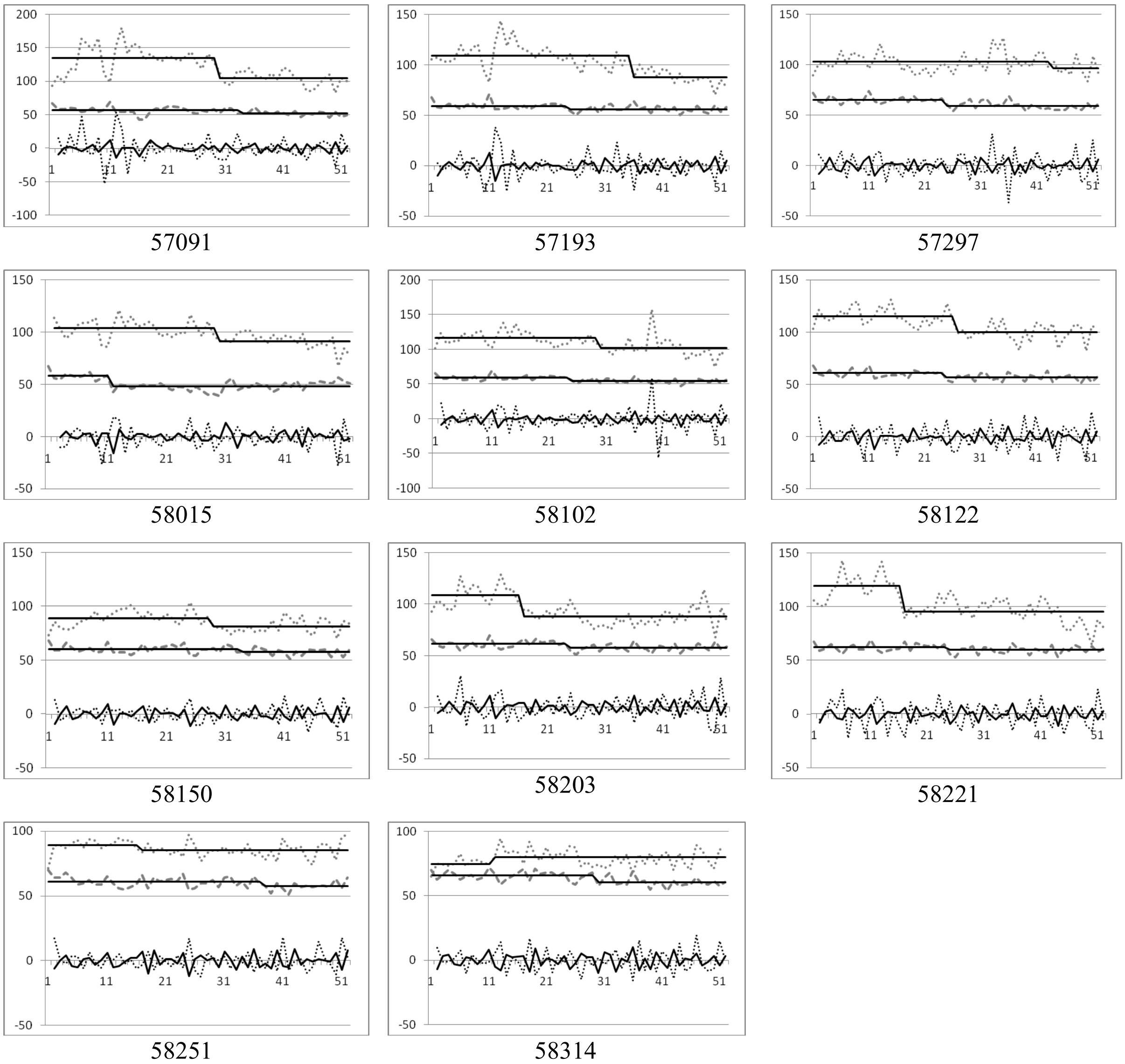

Fig. 4. Annual changes of $E_{pan}$ and $Q$. On the top of every picture: The light gray dotted line represents the annual $E_{pan}$ observation data, the dark gray dashed line represents the annual $Q$ observation data, and the accompanied black polylines represent the average abrupt changes detected via the B–G tests, the turning points represents the most obvious abrupt changes. On the bottom of every picture: The black dotted line fluctuating up and down on the horizontal axis represents d$E_{pan}$/d$t$; and the black solid line represents d$Q$/d$t$. To plot $Q$ and $E_{pan}$ on one coordinate, the horizontal axis is time $t$; for $E_{pan}$, the vertical axis unit is cm/year, and the unit of $Q$ is Okta/year (top part of every picture). For the annual range of change, the vertical axis is d$x$/d$t$, presented as $x_t$-$x_{t-1}$ ('$x$' represents either $Q$ or $E_{pan}$, the unit of d$Q$/d$t$ is Okta/year; the unit of d$E_{pan}$/d$t$ is cm/year; the positive direction represents an increase, and the negative direction represents a decrease) (bottom part of every picture).

The graphs in Fig. 4 show that abrupt changes in $E_{pan}$ and $Q$ do not occur at the same times but do occur sequentially at all stations. All the trends decrease except that for $E_{pan}$ at station 58314. At 7 of the 14 stations (54936, 57089, 57091, 58150, 58203, 58221 and 58251), $Q$ decreases after $E_{pan}$. At 6 of the 14 stations (54916, 57193, 57297, 58015, 58102 and 58122), $Q$ declines before $E_{pan}$. The detrending of the changes in $E_{pan}$ and $Q$ shows that the evaporation paradox is complex and may involve different stages; thus, the relationship between $Q$ and $E_{pan}$ cannot be described as a simple correlation. For convenience, the relationships presented at stations are denoted as “type I”, and the relationships at stations with $E_{pan}$ changes after $Q$ are denoted as “type II”. The station 58314 may present another type of relative change from $E_{pan}$ to $Q$, which can be denoted as “type III”.

Trends of the 51 years and annual changes in $E_{pan}$ and $Q$ are listed in Table 1, Table 2 and Table 3.

| station | nte>0 | nte<0 | Mnte>0 | Mnte<0 | MKTE | MKTE* |
|---|---|---|---|---|---|---|
| 54916 | 21 | 30 | 2120 | -2770 | -0.76 | *** |
| 54936 | 23 | 28 | 1964 | -2306 | -0.56 | *** |
| 57089 | 28 | 23 | 2227 | -2130 | -0.15 | × |
| 57091 | 21 | 30 | 3488 | -3417 | -0.45 | *** |
| 57193 | 22 | 29 | 2370 | -2640 | -0.56 | *** |
| 57297 | 26 | 25 | 2634 | -2622 | -0.09 | ×× |
| 58015 | 26 | 24 | 1770 | -2133 | -0.48 | *** |
| 58102 | 26 | 25 | 3025 | -3140 | -0.49 | *** |
| 58122 | 23 | 28 | 2434 | -2421 | -0.46 | *** |
| 58150 | 23 | 28 | 1661 | -1554 | -0.23 | ** |
| 58203 | 25 | 26 | 2491 | -2585 | -0.36 | *** |
| 58221 | 26 | 25 | 2538 | -2787 | -0.51 | *** |
| 58251 | 26 | 25 | 1598 | -1332 | -0.09 | ×× |
| 58314 | 24 | 27 | 1902 | -1694 | 0.107 | ×× |

Table 1. The trend of $E_{pan}$ in the stations of the Huaihe River Basin between 1954-2005. nte>0: The number of years of $E_{pan}$ increases in the 51 years; nte<0: the number of years of $E_{pan}$ decreases in the 51 years; Mnte>0: the total amount of $E_{pan}$ increases; Mnte<0: the total amount of $E_{pan}$ decreases; MKTE: the total trend of the $E_{pan}$ series by the M-K tests, where ‘-’ indicates a decrease and ‘+’ indicates an increase; MKTE*: significant of the trend, ‘***’ corresponds to $p<0.001$, ‘**’ corresponds to $p<0.05$, ‘×’ corresponds to $p>0.1$, and ‘××’ corresponds to $p>0.25$.

| station | ntc>0 | ntc<0 | ncnt | Mntc>0 | Mntc<0 | MKTC | MKTC* |
|---|---|---|---|---|---|---|---|
| 54916 | 23 | 23 | 5 | 93 | -109 | -0.41 | *** |
| 54936 | 25 | 20 | 6 | 92 | -113 | -0.25 | ** |
| 57089 | 25 | 21 | 5 | 101 | -113 | -0.38 | *** |
| 57091 | 20 | 24 | 7 | 99.22 | -116.2 | -0.38 | *** |
| 57193 | 24 | 22 | 5 | 99 | -109 | -0.28 | ** |
| 57297 | 27 | 22 | 2 | 102 | -112 | -0.44 | *** |
| 58015 | 22 | 25 | 3 | 93 | -98 | -0.14 | × |
| 58102 | 22 | 25 | 4 | 106 | -116 | -0.37 | *** |
| 58122 | 24 | 25 | 2 | 101 | -112 | -0.31 | *** |
| 58150 | 22 | 23 | 6 | 97 | -106 | -0.21 | ** |
| 58203 | 26 | 24 | 1 | 116 | -123 | -0.22 | ** |
| 58221 | 22 | 24 | 5 | 106 | -112 | -0.13 | × |
| 58251 | 24 | 23 | 4 | 100 | -106 | -0.25 | *** |
| 58314 | 27 | 22 | 2 | 101 | -110 | -0.39 | *** |

Table 2. The trend of $Q$ in the stations of the Huaihe River Basin between 1954-2005. ntc>0: The number of years of $Q$ increases in the 51 years; ntc<0: the number of years of $Q$ decreases in the 51 years; ncnt: the number of years of $Q$ has no change in the 51 years; Mntc>0: the total amount of $Q$ increases; Mntc<0: the total amount of $Q$ decreases; MKTC: the total trend of the $Q$ series by the M-K tests, where '-' indicates a decrease and '+' indicates an increase; MKTC*: significant of the trend, '***' corresponds to $p<0.001$, '**' corresponds to $p<0.05$, '×' corresponds to $p>0.1$, and '××' corresponds to $p>0.25$.

| station | necct | necst | station | necct | necst |
|---|---|---|---|---|---|
| 54916 | 27 | 19 | 58102 | 40 | 7 |
| 54936 | 35 | 10 | 58122 | 36 | 13 |
| 57089 | 28 | 18 | 58150 | 32 | 13 |
| 57091 | 29 | 15 | 58203 | 41 | 9 |
| 57193 | 32 | 14 | 58221 | 35 | 11 |
| 57297 | 38 | 11 | 58251 | 33 | 14 |
| 58015 | 29 | 18 | 58314 | 39 | 10 |

Table 3. Comparison on annual changes of $E_{pan}$ and $Q$. necct: The number of years of $E_{pan}$ display an inverse change with $Q$ (e. g., in a certain year, $E_{pan}$ decrease and $Q$ increase) in the 51 years; necst: the number of years of $E_{pan}$ display the same change as $Q$ (e. g., in a certain year, $E_{pan}$ decrease and $Q$ decrease too).

At all the stations, $E_{pan}$ presented more inverse annual changes than $Q$ did. Among the 14 stations, 7 displayed more years with increases in $Q$ (ntc>0) than with decreases (ntc<0) out of the 51 annual changes; additionally, 6 displayed more years with decreases than with increases, and one had equal numbers of years with increases and decreases. However, whether the difference in the number of years $Q$ increased or decreased varied between 1 year and 5 years. At all the stations, the total number of years with a decrease in $Q$ (Mntc<0) was slightly greater than the total number of years with an increase in $Q$ (Mntc>0). For $E_{pan}$, 5 of the 14 stations displayed slightly more years with an increase (nte>0) than with a decrease (nte<0), and 3 of the 5 stations displayed no obvious trends in $E_{pan}$. At all the stations, the difference in the number of years $E_{pan}$ increased or decreased ranged from 1 year to 9 years. At 6 stations, the total $E_{pan}$ increase in corresponding years (Mnte>0) was greater than the total $E_{pan}$ decrease in years with the opposite trend (Mnte<0); all 4 stations with negligible trends (p>0.1 and p>0.25) in $E_{pan}$ were included among these 6 stations. The data suggest that at small time scales (annually), $Q$ and $E_{pan}$ do not display contrary trends but are characterized by some uncertainty, as $E_{pan}$ presents the same annual trends as $Q$ does in some years.

*b). Oscillations in the cloud quantity and $E_{pan}$*

The data from the Huaihe River Basin show that at the annual time scale, $Q$ and $E_{pan}$ mainly present contrary trends. The long-term trends, abrupt changes and relative changes at the annual scale show that in this region, at the annual scale, $E_{pan}$ generally decreases as $Q$ increases. However, the long-term changes do not follow this trend; rather, a sequential decreasing trend generally exists.

Different annual changes in $E_{pan}$ and $Q$ occur, as these variables are both sensitive to small changes in the other. In the long term, both $E_{pan}$ and $Q$ display decreasing trends with annual uncertainty. This phenomenon cannot be described via linear mathematics.

The relationships between $E_{pan}$ and $Q$ in terms of their annual changes can be summarized as follows: 1) the decadal changes in $Q$ and $E_{pan}$ display some oscillatory properties, and the changes may have various stages and steps. 2) The general trends of $E_{pan}$ and $Q$ in the Huaihe River Basin are decreasing, but in terms of annual changes, the changes in these variables are often contrasting; thus, at different time scales, they present different correlations. 3) The abrupt change test show that $E_{pan}$ and $Q$ change forms of the stations in the Huaihe River Basin can be classified into three different types Thus, obvious uncertainty and delay between $Q$ changes and $E_{pan}$ changes exist in the system, and the interactions between $E_{pan}$ and $Q$ could affect the overall $Q$ and $E_{pan}$ trends.

On the basis of the above analyses, a nonlinear second-order neutral delay dynamic equation (the NSNDE), which is a part of the Duffing equations (Liu et al. 2003; Loria et al. 1998), is employed here to illustrate the system. The following set of dynamic differential equations is used to model a periodic or quasiperiodic system with different relationships at different time scales and with neutral delays in periods with driving forces and response (in this system, both $Q$ and $E_{pan}$ are considered driving forces) (an der Heiden et al. 1990; Han et al. 2009).

For long-term series, both $E_{pan}$ and $Q$ display multiple waves and periods, and the function of $E_{pan}$ or $Q$ can be written as $y = F(x)$, where x represents $E_{pan}$ or $Q$. Let $x \propto F(x)$ for a given fluctuation in $F(x)$ shift in the same direction as $E_{pan}$ or $Q$. $y$ is also a function of time $t$. The system of $E_{pan}$ or $Q$ changes with time can be presented as an NSNDE:

$$\ddot{y} + a_1\dot{y} + a_0 y = A\sin\omega(t-\tau) \quad (11)$$

where $\tau$ is a time delay, $\omega$ is the frequency of the driving force and $A$ is its maximum amplitude. The derivatives are obtained with respect to $t$. In the equation, the right-hand side is referred to as the delayed restoration force, and the items on the left-hand side represent the acceleration, damping force and restoration force, respectively.

Over time, eq. (11) has a solution in the form of

$$y = B \sin \omega_1 t \tag{12}$$

where $B$ is the amplitude of the dependent variable and its frequency is $\omega_1$. For $\omega$, the input frequency, $\omega_1$ is the output frequency; they are both time variables, but here, the subscript is neglected.

$F(Q)$, the driving force, and $F(E_{pan})$, the response, at the same station and with the same phase at time 0 are plotted on the same coordinates, as shown in Fig. 5. Over time, the angular amplitude and period of the response increase until they have opposite phases, and they cannot be in the same phase at the same time $t$ again.

From this simple conceptual model (in which dimension units and minimum time units are neglected), the three types of evaporation changes related to cloud quantity can be described as three different stages of long-term large-scale processing (as shown in Fig. 5), which can be repeated over time. Fig. 5 also shows that in the same region, different stages may correspond to different relationships between $E_{pan}$ and $Q$ because they increase and decrease with time, and the relationships, which are classified as three types, repeat but are not exactly the same as they were at previous times.

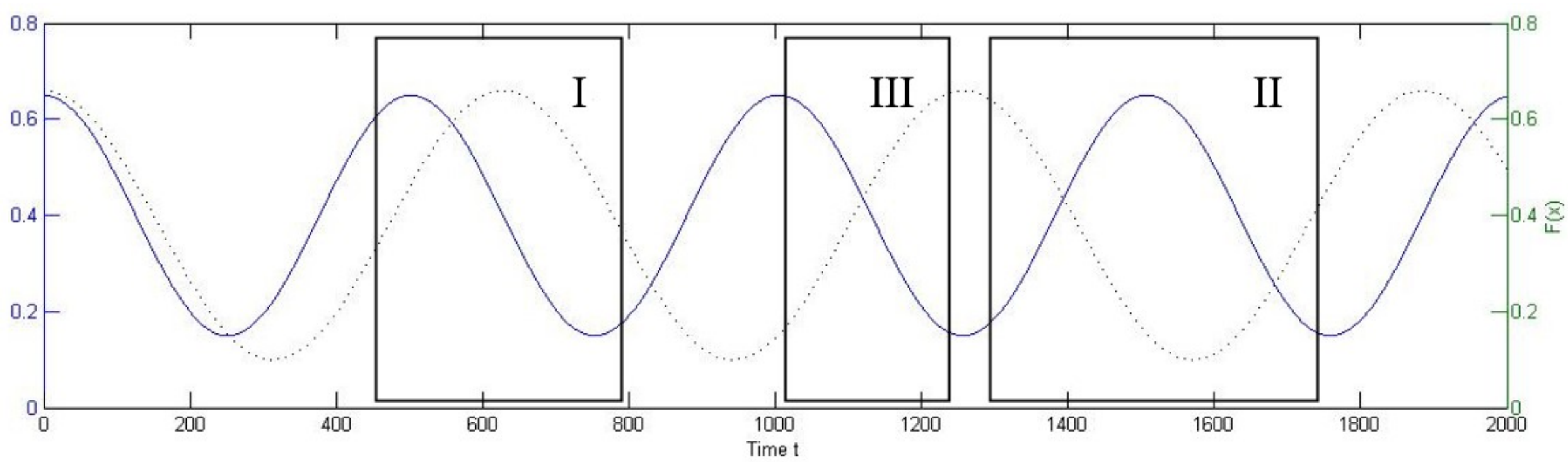

Fig. 5 The $x$-axis represents time $t$, and the initial point can be any time ($E_{pan}$ and $Q$ are exactly in the same phase). The $y$-axis represents trigonometric functions of t in the form of eq. (2). The waveform of the solid line represents $F(Q)$, whereas the dashed line represents $F(E_{pan})$. Sections 'I' and 'II' present two types of evaporation paradoxes observed in the Huaihe River Basin, and 'III' presents a decrease in $E_{pan}$ with increasing $Q$.

## 4. Theory description and the mathematical expression

*a) Description on the steamer model for the relationship between $E_{pan}$ and cloud quantity*

The above analysis of data from the Huaihe River Basin reveals that in this region, $E_{pan}$ and cloud quantity present different relationships at different time scales, but both fluctuate with uncertainty; thus, the system presents characteristics of chaos, which may be caused by the mixed effects of aerosols, clouds, and water cycling; notably, these factors constitute a dynamic system. In this section, the relationship between changes in the cloud quantity and $E_{pan}$ with time is written as an NSNDE.

Aerosols are the product of physical and chemical processes involving atmospheric particles, which are important resources of cloud condensation nuclei (CCN). Increasing cloud quantities are related to increasing CCN, and water vapor evaporates into the atmosphere and accumulates on aerosol granules, generating cloud droplets. In the context of increasing temperature and increasing number of particles being released into the atmosphere, evaporation from the surface of water bodies increases. In a short time, $E_{pan}$ may display an increasing trend, and consequently, cloud drops are generated quickly; thus, the quantity of clouds increases after $E_{pan}$ increases. This process continues, and the cumulative cloud body begins to form a 'steamer' (Chinese phonetic alphabet: Zheng Long, always made from bamboo, is a cooking tool that heats food with steam). Vapor begins to evaporate, and cloud droplets do not readily form; however, vapor is retained by the steamer, although CCN release continues as the pressure and temperature increase. The humidity of the atmosphere increases steadily during this stage, and humid surroundings inhibit the evaporation of open-channel water; therefore, $E_{pan}$ generally begins to decrease during this stage and presents a phenomenon called the 'evaporation paradox'. Thus, this stage is referred to as the "formation of the paradox". In this stage, $E_{pan}$ decreases as the cloud quantity continues to increase. This is what happen in the regions show the evaporation paradox exist with cloud quantity increase.

In stage when $E_{pan}$ decreases or stabilizes, particulate matter is released into the atmosphere. The increase in CCN and decrease in particulate volume (as more small particulate matter is released into the atmosphere) cause water molecule accumulation to decrease, with insufficient evaporation of water vapor. Additionally, some CCN cannot form cloud droplets or generate clouds but remain in the air. Thus, clouds become scattered and thinner, the ‘steamer’ structure dissipates but does not disappear. Moreover, the humidity of the atmosphere is still high, and there is still considerable free water vapor and many small aerosols in the atmosphere, which promote water evaporation; therefore, $E_{pan}$ continues to decrease. In this stage, both the cloud quantity and $E_{pan}$ decrease; here, it is called the “duration of the paradox”. The type I paradox observed in the Huaihe River Basin is a representation of this stage. This is what happen in part of the regions where evaporation paradox exists but cannot be interpreted by cloud quantity increase.

In the process of cloud scattering, the amount of shortwave radiation from the sun that reaches the Earth’s surface increases, and the temperature continues to increase. Because clouds remain scattered and shortwave radiation increases, in the next stage, $E_{pan}$ begins to increase, and the evaporation paradox no longer holds; this phenomenon can be called the “waning of the paradox”. If the amount of small particulate matter continues to increase, the evaporated water that accumulates on particles cannot generate clouds but remains in the atmosphere, resulting in foggy weather, which increases the relative humidity. Alternatively, a decrease in the quantity of cloud droplets decreases the relative humidity. If $E_{pan}$ decreases again and $E_{pan}$ decreases after the cloud quantity decreases, the type II paradox occurs, described as the “reappearance of the paradox”. In some cases, $E_{pan}$ may just reach a valley and then increase, whereas the cloud quantity may peak and then decrease. Thus, the cloud quantity decreases after $E_{pan}$ increases, as observed at station 58314 (type III).

The evaporation paradox, which intensifies and then weakens, may constitute a repetitive cycle intensified by human activities, e.g., industry and transpiration may release more artificial particles into atmosphere. The disequilibrium of global evaporation and clouds may be affected by variations in terrain, atmospheric circulation and imbalances in atmospheric particle release. In different environments, different evaporate types (type I, type II and type III expounded above) may exist, and different stages of the evaporation paradox may be observed. The speed of release, amount and type of particulate matter may influence the period of each stage of the paradox.

The steamer model illustrates the correlation established on the basis of global cloud quantity and evaporation distribution observations and encompasses the mechanism of how cloud quantity affects $E_{pan}$. Because the model has a basic structure and includes turbulence, the “steamer” can be represented by a set of chaos-based dynamic system equations. The cloud quantity and evapotranspiration constitute a linked system modeled in the context of increases in temperature and particle emissions.

*b). Mathematical expression of the theory*

In the above analysis, an NSNDE was used to assess the interactions between $E_{pan}$ and the quantity of clouds, either of the variables can be driving force of the relation or response to changes in another. A simplified NSNDE can be combined with a hydrological equation to quantitatively express the evaporation paradox. By integrating eq. (11) with x and calculating the transpose of the result, we obtain

$$y = x - a_1 \int y dx + a_0 \iint y dx - \frac{A}{2} x^2 \sin \omega (t - \tau) \tag{13}$$

Because this formula is a periodic function, it can be expressed as

$$y(x,t) = y\left(x, t + \frac{2\pi}{\omega}\right) \tag{14}$$

This formula yields a Poincare map (Hirsch et al. 2008) with a periodicity of $y$; on this basis, the problem can be simplified to a 1-D discrete dynamic system, and a simple iterative function can be used to present it in ideal form. The most popular and effective iterative function for dynamic systems is the logistic map (Hirsch et al. 2008; Liu et al. 2003; Lorenz 1963), which has been used in weather forecasting (Lorenz 1963), climate change research (Rial 2004), earthquake research (Sykes et al. 1999) and ecosystem analysis (Odenbaugh 2005) to depict series of random events with periodic properties (Liu et al. 2003).

The "steamer" model starts at the link between the logistic map and the well-known Priestley-Taylor equation (1972). For evaporation, a logistic map (Hirsch et al. 2008; Liu et al. 2003) is used to describe the relationship between $E_{pan}$ at time $t+1$ and $t$; here, suppose that the value of $2\pi/\omega$ in eq. (14) is the "minimum unit" in the new function.

Let $F(E_{pan}) = E = \frac{E_{pan}}{E_0}$, and $E_0$ be the maximum potential evaporation in the region; thus, $E_{pan} \propto E$ can be expressed as an iterative function.

A logistic map-like form of this relation can be written as

$$E_{t+1} = \alpha E_t - \beta E_t \tag{15}$$

In this case, $E_t$ represents $E$ at time $t$, $\alpha$ is a driving factor, and $\beta$ is a consumption factor. From this, the stability can be determined as $E_{t+1} = E_t$ when

$$E_t = \frac{\alpha - 1}{\beta} \tag{16}$$

The evaporation paradox occurs when $E_t < \frac{\alpha - 1}{\beta}$ and $E_{t+1} < E_t$. In addition, evaporation increases with increasing temperature.

The Priestley–Taylor equation is as follows:

$$\lambda E_{pan} \approx 1.26\left(\frac{s}{s+\gamma}\right)R_n \tag{17}$$

where $\lambda$ (~2.4 MJ kg-1) is the latent heat of vaporization of water, $R_n$ (J m-2 s-1) is the net irradiance, γ (~67 Pa K-1) is the psychrometric constant, and $s$ is the slope of the saturation vapor pressure–temperature relationship at temperature $T$. From observation in eq. (16) and eq. (17) and obtain the follows:

$$\alpha = 1.26 s R_n + 1 \tag{18}$$

and

$$\beta = \lambda E_0 \left(s + \gamma\right) \tag{19}$$

Suppose that when $E_{t+1} = E_t$, $R_n = R_n^0$. Thus, when $E_{t+1} < E_t$,

$$\frac{\alpha - 1}{\beta} > \frac{1.26 s R_n}{\lambda E_0 \left(s + \gamma\right)} \tag{20}$$

for $R_n < R_n^0$.

Assuming that $E_t^{'} = \frac{\alpha}{\beta} E_t$, the following expression can be obtained

$$E_{t+1} = \frac{\alpha^2}{\beta} E_t^{'}\left(1 - E_t^{'}\right) \tag{21}$$

The mapping process is sensitive to system dynamics in the following ways (Hirsch et al. 2008; Liu et al. 2003): 1) the process is very sensitive to the initial conditions for $E$ and varies with minor environmental changes, 2) for any value of $E$, when $t$ continuously varies, some points on the track of $E_t$ must be infinitely close to the original track; and 3) the mapping is continuous with respect to t. Denoting $\mu = \frac{\alpha^2}{\beta}$, in a system with chaos, this parameter is in the range of [3.57, 4] (Lorenz, 1963), and $\mu$ increases as chaos intensifies. Additionally, in eq. (21), if $\mu$ increases, $E_{t+1}$ increases; thus, $E_{t+1} > E_t$. Then, eq. (20) can be rewritten such that $R_n > R_n^0$, which means that the cloud quantity decreases and $R_n$ increases.

Now, the lag effect of the change in $E_{pan}$ is described by a nonlinear map. Given that $R_n$ decreases as $Q$ increases, we establish the following relationship:

$$R_n = \frac{k}{Q} \tag{22}$$

where $k$ is a constant.

Substituting eq. (16), eq. (17) and eq. (19) into eq. (21) yields

$$E_{t+1} = \frac{(1.26sk)^2}{\lambda E_0 Q_t^2 (s+\gamma)} E_t^{'} \left(1 - E_t^{'}\right) \tag{23}$$

where $Q_t$ is the cloud quantity at time $t$. This equation reflects the reverse trend relationship between $Q$ and $E$. $E_{t+1}$ decreases as $Q_t$ increases.

In addition, $Q_t$ can be expressed in a logistic mapping-like form as

$$Q_t = \mu_2 Q_{t-1}\left(1 - Q_{t-1}\right) \tag{24}$$

This expression can be substituted into eq. (23) to obtain

$$E_{t+1} = \frac{(1.26sk)^2}{\lambda E_0 \mu_2^2 Q_{t-1}^2 (1-Q_{t-1})^2 (s+\gamma)} E_t^{'} \left(1 - E_t^{'}\right) \tag{25}$$

Eq. (25) can be written in short notation as

$$E_{t+1} = f\left(E_t, Q_{t-1}\right) \tag{26}$$

which presents the mapping relation

$$Q_{t-1} \rightarrow E_{t+1} \tag{27}$$

The mapping expression indicates that a change in $Q$ is a driver of changes in $E$ and that a change in $E_{pan}$ lags behind changes in $Q$.

Additionally, as the inverse function of eq. (25), the function for relating $E$ to $Q$ is as follows:

$$Q_t = \frac{1}{2}\left\{\sqrt{\frac{1.26sk}{\mu_2}}\left[\frac{E_{t+1}^{'}\left(1 - E_{t+1}^{'}\right)}{\lambda E_0 E_{t+2}}\right]^{\frac{1}{4}} + 1\right\} \tag{28}$$

The periodicity of $E$ and $Q$ proven in eqs. (13) and (14) can be written as

$$Q_t = \frac{1}{2}\left\{\sqrt{\frac{1.26sk}{\mu_2}}\left[\frac{E_{t-\omega\pi+1}^{'}\left(1 - E_{t-\omega\pi+1}^{'}\right)}{\lambda E_0 E_{t-\omega\pi+2}}\right]^{\frac{1}{4}} + 1\right\}, \quad \omega = 1, 2, \ldots \tag{29}$$

$$\because \quad t - \omega\pi + 2 < t$$

$$\therefore \quad E_{t-\Delta t} \rightarrow Q_t \quad (\text{denote } \Delta t = \omega\pi\text{-}2) \tag{30}$$

This relationship indicates that evaporation can affect the cloud quantity just as it is affected by the cloud quantity and has a peak or valley phase before that of the cloud quantity. Relations (27) and (30) indicate that the cloud quantity and $E_{pan}$ compose a dynamically linked system with different phases under different conditions.

## 5. Discussion

*a). Relationship between actual evaporation ($E_a$) and cloud quantity*

$E_{pan}$ reflects actual environmental evaporation ($E_a$) but not exactly equal to it. Their complementary relationship (Brutsaer and Parlanger 1998) explains the evaporation paradox, although it is not suitable for paradoxes in all regions. A theory for the relationship between clouds and evaporation can be built on the basis of the "steamer" model.

*a) Analysis of sensible heat flux change*

The complementary relationship indicates below relation exists

$$E_a = kE_{pan} - \Delta H \tag{31}$$

where $\Delta H$ represents the increase in sensible heat flux.

According to the explanation of the complementary relationship to evaporation paradox, a decrease in $E_{pan}$ means that $E_a$ increases, so $\Delta H$ should decrease. The sensible heat flux $H$ (Businger, 1975) can be expressed as

$$H = \rho C_p C_D U \left(T_0 - T\right) \tag{32}$$

where $\rho$ is the atmospheric density, $C_p$ is the specific heat of air, $C_D$ is the transfer coefficient for the sensible heat flux, $U$ is the wind speed at a height of 2 m above the surface, $T_0$ is the surface temperature, and $T$ is the air temperature.

We consider a short time period dt. In this period, although $\rho$, $C_p$ and $C_D$ depend on $U$, $T$ and other meteorological factors, parametric methods are used to determine them; thus, every specific value of the factors can be considered a constant for simplicity. $H$ can be differentiated with respect to t to obtain

$$\Delta H=\frac{dH}{dt}=\rho C_p C_D\frac{d\left[U\left(T_0-T\right)\right]}{dt}=\rho C_p C_D\left[U\frac{d\left(T_0-T\right)}{dt}+\left(T_0-T\right)\frac{dU}{dt}\right]=\rho C_p C_D\left[U\left(\frac{dT_0}{dt}-\frac{dT}{dt}\right)+\left(T_0-T\right)\frac{dU}{dt}\right] \tag{33}$$

The differential terms can also be written as $\Delta T_0$, $\Delta T$ and $\Delta U$, so the parameters are separated in dt. As time t elapses, if $\Delta H$ decreases, for a specific parameter, $\Delta T_0$ and $\Delta U$ should decrease and $\Delta T$ should increase, or if $T_0$, $T$ and $U$ are the parameters, $T_0$ and $U$ should decrease and $T$ should increase. This means that the relationships between sensible heat flux and the three meteorological factors are the same over a specific short time and long period. In the context of a universal temperature increase and a decrease in wind speed, these conditions are satisfied for $T$ and $U$ in many regions. However, for $T_0$, the effects of clouds should be considered, which is a way to connect the separate theories for cloud quantity vs $E_{pan}$ and $E_{pan}$ vs $E_a$.

G.W. Paltridge (1974) reported that the cloud quantity and surface temperature $T_0$ satisfy the requirements of the global energy balance, which means that the absorbed solar radiation must equalize the loss of long wave radiant energy to space. When the effects of cloud height and other cloud characteristics are discarded, a 1% increase in the solar constant $I_0$ will cause a 0.009 change in fractional cloud cover or a 0.35 K change in $T_0$; additionally, an increase in surface albedo (which may be related to land cover) can decrease both $T_0$ and the quantity of clouds. An increase in cloud reflectivity can lower $T_0$ and the quantity of clouds, which means that if the clouds and sky become "brighter", it will lower $T_0$ and the quantity of clouds. Aerosols in the lower atmosphere generally reduce the cloud quantity but increase $T_0$. Qian and Huang (1990) noted that when other factors are discarded and only the cloud quantity and $T_0$ are considered, although that clouds have both cooling and greenhouse effects on surface temperature, when total the cloud quantity increases, $T_0$ decreases. When $I_0$ varies by less than 1% annually, its effect on $T_0$ and the cloud quantity can be neglected. Moreover, as land cover changes from rough to smooth, e.g., from woodland to tarmacs or car parks, surface albedo increases, and in these regions, $T_0$ may increase. Universal darkening trends exist in the Asian atmosphere, and clouds, wide coal fields and heavy industrial areas are common in the Huaihe River Basin; moreover, there is no evidence to support whether clouds become lighter, although the quantity of clouds decreases. When all other factors are discarded, an increase in the cloud quantity reduces $T_0$, so it is very likely that the complementary relationship is not valid in most parts of the Huaihe River Basin, except in cases when $T_0$ decreases as the cloud quantity decreases.

*b). The oscillation of $E_a$ is related to the cloud quantity and $E_{pan}$ change*

In 'steamer' theory, $E_a$ may fluctuate with time as the cloud quantity and $E_{pan}$ fluctuate. In the formation stage of the paradox, $E_{pan}$ begins to decrease with increasing cloud quantity, $T_0$ tends to decrease, and $\Delta H$ tends to decrease, so $E_a$ displays a complementary relationship with $E_{pan}$. In the duration stage of the paradox, as clouds become scattered and thinner but evaporated water remains in the lower atmosphere, $E_a$ is likely to decrease; however, according to the complementary relationship, $E_a$ increases as $E_{pan}$ decreases. Therefore, in this stage, the complementary relationship does not hold in the context of the evaporation paradox. Then, in the dissipation stage of the paradox, the cloud quantity continues to decrease, and $E_{pan}$ begins to increase. Changes in $E_a$ are mainly dependent on the amounts and types of aerosols and the surface characteristics. If the effect of the decrease in cloud quantity on $E_a$ exceeds the effects of other factors, $E_a$ decreases as expected. When the paradox reappears, $E_{pan}$ decreases after the cloud quantity decreases; because the cloud quantity results in a longer duration of decrease than the $E_{pan}$ and aerosol effects do, $E_a$ is more likely to decrease than to increase, and consequently, in this stage, the evapotranspiration conditions may not satisfy the complementary relationship. Thus, the complementary relationship meets or not varies with the stages of the evaporation paradox, such as waning or reappearance stage; it holds in many cases, but not always and everywhere. The cloud quantity affects when and where it holds, considering the influence of other factors, such as aerosols and surface characteristics.

*c). Discussion on energy generation*

In dynamic systems, Lyapunov functions (Hirsch 2008) are commonly used to analyze the stability of the system; in systems based on mechanics, electricity and matter (Pykh 2003), energy is usually present and transferred. Entropy is reflected by energy generation per unit time in dynamic system theory (Pykh 2001; Hangos 2010); if a Lyapunov function is present in entropy form in a system, positive entropy generation results in the system becoming stable, and negative entropy generation means results in it becoming unstable (Hangos 2010). The transfer of momentum and energy accompanied by water cycling drives the system to change, and the system may be included by various weather phenomena. Thus, two Lyapunov functions can be used to express entropy generation in the processes of evaporation generation and dissipation; here, they are applied to assess the corresponding conditions of the atmosphere and land surface.

For the land surface, the net energy income is

$$G_g = G_T + R_n + G_r - E \tag{34}$$

Here, entropy generation per unit time is denoted as $\frac{d(\Delta G_g)}{dt}$. For the atmosphere, the net energy income is

$$G_A = G_T - R_n - G_r + E \tag{35}$$

and entropy generation per unit time is expressed as $\frac{d(\Delta G_A)}{dt}$. In eqs. (34) and (35), $G_T$ is the energy associated with the temperature increase, and $G_r$ is the energy associated with precipitation. Here, we consider entropy generation linked to changes in net longwave radiation ($R_n$) and evaporation ($E$) only.

In the formation stage of the evaporation paradox, $\frac{dR_n}{dt} < 0$, $\frac{dE}{dt} < 0$, and $\left|\frac{dR_n}{dt}\right| > \left|\frac{dE}{dt}\right|$, and thus, in this stage, $\frac{d(\Delta G_g)}{dt} > 0$, which means that the conditions of the land surface getting stable. When $\frac{d(\Delta G_A)}{dt} < 0$, the conditions of the atmosphere getting unstable.

In the duration stage of the evaporation paradox, $\frac{dR_n}{dt} < 0$, $\frac{dE}{dt} > 0$, and $\left|\frac{dR_n}{dt}\right| > \left|\frac{dE}{dt}\right|$; therefore, $\frac{d(\Delta G_g)}{dt} < 0$ and $\frac{d(\Delta G_A)}{dt} > 0$ indicate that the land surface getting unstable and that the atmosphere getting stable.

In the waning stage of the evaporation paradox, $\frac{dR_n}{dt} > 0$, $\frac{dE}{dt} > 0$, and $\left|\frac{dR_n}{dt}\right| < \left|\frac{dE}{dt}\right|$; therefore, $\frac{d(\Delta G_g)}{dt} < 0$ and $\frac{d(\Delta G_A)}{dt} > 0$ indicate that the conditions at the land surface getting unstable and that the atmosphere getting stable.

In the recurring stage of the evaporation paradox, $\frac{dR_n}{dt} > 0$, $\frac{dE}{dt} < 0$, and $\left|\frac{dR_n}{dt}\right| < \left|\frac{dE}{dt}\right|$; therefore, $\frac{d(\Delta G_g)}{dt} > 0$ and $\frac{d(\Delta G_A)}{dt} < 0$ indicate that the conditions of the land surface getting stable and those of atmosphere getting unstable.

In summary, in the stages in which the evaporation paradox emerges and reappears and the atmosphere becomes unstable, chaos intensifies, which may be attributed to the intensified occurrence of extreme weather events in systems with high chaos and energy release; these systems, such as those associated with extreme precipitation and drought, can have many states (Lorenz 1963). On the other hand, in the duration and waning stages of the evaporation paradox, the land surface becomes unstable, which may cause sensible and latent fluxes to increase and result in warming, which may increase land temperature differentiation.

## 6. Conclusion

To reveal how clouds affect $E_{pan}$, cloud quantity and $E_{pan}$ records from 1954—2005 in the Huaihe River Basin were investigated. During this 52-year period, the cloud quantity in the Huaihe River Basin changed from decreasing slightly to increasing slightly or remained steady in most cases; however, at some stations, it gradually decreased. $E_{pan}$ presented similar trends as the cloud quantity in this period but with different phases. In addition, annually, the $E_{pan}$ trend displays an inverse direction as the cloud quantity trend; in other words, it decreases when the cloud quantity increases, and vice versa. The abrupt change points of the cloud quantity and $E_{pan}$ series differ, but the intervals of change are similar, suggesting that the changes in the cloud quantity and $E_{pan}$ do not occur at the same time and are lagged. In terms of annual changes, although $E_{pan}$ generally increases when the cloud amount decreases, it sometimes decreases, too.

Alternative increases and decreases in the cloud quantity and $E_{pan}$ and the time intervals of the abrupt changes in the average totals show that both display cyclic characteristics; they affect each other but do not change at the same time, and their intervals of change vary. Thus, an NSNDE was used to describe the long-term oscillations in the cloud quantity. Moreover, series of the $E_{pan}$ displays wave-like form with distinct changes in frequency and wavelength. This pattern reflects different evaporation characteristics worldwide, encompassing both decreases in $E_{pan}$ while the cloud quantity increases and decreases.

The ‘steamer’ model, differing from the present theories used to explain evaporation distributions with changes in cloud quantity, was developed with the NSNDE of $E_{pan}$ considering long-term changes in cloud quantity. The evaporation paradox can be divided into stages: when the background temperature increases and more aerosol granules are released into the atmosphere, more vapor evaporates from water surfaces. In a short period, $E_{pan}$ may increase; then, within a short time, with the generation of cloud droplets, the quantity of clouds increases. This process continues, and the clouds work as a ‘steamer’. $E_{pan}$ decreases after the quantity of clouds increases, which is referred to as ‘the formation of the paradox’, and reflects the situation observed in many regions of the world. In the $E_{pan}$ decreasing and maintenance stage, the clouds become scattered and thinner. As the quantity of clouds decreases, the ‘steamer’ structure broken but does not disappear, and there is still much free water vapor in the atmosphere to suppress water evaporation; thus, $E_{pan}$ continues to decrease. In this stage, both the cloud quantity and $E_{pan}$ decrease, which is referred to as ‘the duration of the paradox’. This situation is observed at half of the national stations in the Huaihe River Basin. Then, as the quantity of clouds decreases, $E_{pan}$ increases, which is referred to as ‘the waning of the paradox’. If small CCN continue to increase, the evaporated water accumulated on the CCN tend to remains in the atmosphere than to generate clouds, and create surround with increasing relative humidity and inhibits evaporation. Therefore, in this situation, $E_{pan}$ decreases, and without a sufficient quantity of cloud droplets, the quantity of clouds continues to decrease; thus, in this stage, $E_{pan}$ decreases with decreasing cloud quantity, which is referred to as ‘the recurring of the paradox’; notably, this stage is observed at 6 of the 14 stations in the Huaihe River Basin. After $E_{pan}$ and the quantity of clouds decrease and then increase, on the basis of the NSNDE, the wave-like series of the $E_{pan}$ may reach a valley and then increase when the series of the cloud quantity reach a peak and then decrease, as observed at station 58314.

The quantity of clouds and evapotranspiration are linked in the current system of increasing temperatures and high emissions of CCN, and the ‘evaporation paradox’ varies across stages and over time. The increases and decreases in cloud quantity and evaporation display an inverse relationship. The relationship between actual evaporation $E_a$ and $E_{pan}$ is not always complementary and depends on many factors, including the cloud quantity and others. An analysis of the increase in the sensible heat flux indicated $E_a$ has similar relationships with the cloud quantity and $E_{pan}$; it fluctuates with time, and its relationship with the cloud quantity cannot be illustrated by a simple correlation.

The present theories used to explain the paradox are part of a dynamic system relating cloud quantity and evaporation. Clouds may impact evaporation directly and indirectly through controlling humidity and radiation and constrain the spatial and temporal distributions of precipitation through accumulation and scattering processes. The relation between the land surface and the atmosphere reflects global energy transfer. The stability of atmosphere and land surface alternating in the process of energy transpiration, that would make chaos in the Earth system intensify and cause more extreme events in some periods but this situation may alleviate in other periods, and frequency of the extreme, e. g., extreme precipitation, high or low temperature, draught or flood, and the interval between a kind of weather event (e. g., less precipitation) and another (e. g., draught) associate with the level of chaos.

*Acknowledgments*

This work is supported by the Open Research Fund Program of the State Key Laboratory of Hydroscience Science and Engineering (Application of a theorized delayed response dynamic model in the construction of a digital 3-D river network, sklhse-2024-B-01) and the National Fund Cultivation Program of Jimei University (Study on the impact of spatial emissions from maritime traffic on air–sea interactions and its medium-term climate response, ZP2023005). These organizations provide open access to data.

*Data availability statement*

The data that support the findings of this study are openly provided by the national Center for Atmospheric Research (NCAR) at https://oidc.rda.ucar.edu/datasets/d292002/#, the International Satellite Cloud Climatology Project (ISCCP) at https://isccp.giss.nasa.gov/products/onlineData.html, and the China Meteorological Administration at http://www.cdc.cma.gov.cn/.

REFERENCES

an der Heiden, U., A. Longtin, M. C. Mackey, J. G. Milton, and R. Scholl, 1990: Oscillatory modes in a nonlinear second-order differential equation with delay. *Journal Dyn Diff Equat* ., **2**, 423-449, https://doi.org/10.1007/BF01054042.

Arking, A., 1991: The radiative effects of clouds and their impact on climate. *Bull. Amer. Meteor. Soc.*, **72**, 795-813, https://doi.org/10.1175/1520-0477(1991)072<0795:TREOCA>2.0.CO;2.

Bernaola-Galván, P., P. Ch.Ivanov, L. A. N. Amaral, and H. E. Stanly, 2001: Scale invariance in the nonstationarity of human heart rate. *Phys. Rev. Lett.*, **87**: 168105-1-4, 10.1103/PhysRevLett87.168105.

Brutsaer, W., and M. B. Parlanger, 1998: Hydrologic cycle explains the evaporation paradox. *Nature*., **396**, 30, https://doi.org/10.1038/23845.

Burn, D. H., and N. M. Hesch, 2007: Trends in evaporation for the Canadian Prairies. *J. Hydrology.*, **336**, 61-73, https://doi.org/10.1016/j.jhydrol.2006.12.011.

Businger, J. A., 1975: Interactions of sea and atmosphere. *Rev. Geophys.,* **13**, 720-822. https://doi.org/10.1029/RG013i003p00720.

Chattopadhyay, N., and M. Hulme, 1997: Evaporation and potential evapotranspiration in India under conditions of recent and future climate change. *Agric. For. Meteorol.*, **87**, 55-73, https://doi.org/10.1016/S0168-1923(97)00006-3.

Cong Z.-T, D.-W. Yang, and G.-H. Ni, 2009: Does evaporation paradox exist in China? *Hydrol. Earth Syst. Sci.*, **13**, 357-366, https://doi.org/10.5194/hess-13-357-2009.

Ding S.-G, G.-Y. Shi, and C.-S. Zhao, 2004: Analyzing global trends of different cloud types and their potential impacts on climate by using the ISCCP D2 dataset. *Chin. Sci. Bull.*, **49**, 1301-1306, https://doi.org/10.1360/03wd0614.

Farquhar, G. D., and M. L. Roderick, 2003: Pinatubo, diffuse light, and the carbon cycle. *Science*., **299**, 1997-1998, https://doi.org/10.1126/science.1080681.

Golubev, V. S., A. N. Gorshkov, S. N. Mokhov, A. V. Blyakharchuk, V. V. Borisov, V. P. Gruza, E. G. Dvoynikov, and T. C. Peterson, 2001: Evaporation changes over the contiguous United States and the former USSR: A reassessment. *Geophys. Res. Lett.*, **28**, 2665-2668, https://doi.org/10.1029/2000GL012851.

Hahn, C. J., and S. G. Warren, 2009: Extended Edited Cloud Reports from Ships and Land Stations over the Globe, 1952-1996 (2009 update). Carbon Dioxide Information Analysis Center Numerical Data Package NDP-026C, 79pp, d292002 | DOI: 10.5065/VA78-8E98

Han, Z. L., T.-X. Li, S.-R. Sun, and C.-H. Zhang, 2009: Oscillation for second-order nonlinear delay dynamic equations on time scales. *Adv. Differ. Equ.*, 756171, https://doi.org/10.1155/2009/756171.

Hangos, K. M., 2010: Engineering model reduction and entropy-based Lyapunov functions in chemical reaction kinetics. *Entropy*., **12**, 772-797, https://doi.org/10.3390/e12040772.

Hirsch, M. W., S. Smale, R. L. Devaney, 2008: *Differential Equations, Dynamical Systems, and Introduction to Chaos Posts & Telecom*. Academic Press, 425 pp.

Hobbins, M. T., J. A. Ramírez, and T. C. Brown, 2004: Trends in pan evaporation and actual evapotranspiration across the conterminous U.S.: Paradoxical or complementary? *Geophys. Res. Lett.*, **31**, L13503, https://doi.org/10.1029/2004GL19846.

Jiang, H. L., and G. Feingold, 2006: Effect of aerosol on warm convective clouds: Aerosol-cloud-surface flux feedbacks in a new coupled large eddy model. *J. Geophys. Res.*, **111**, D01202, https://doi.org/10.1029/2005JD006138.

Kendall, M., and J. D. Gibbons, 1990: *Rank Correlation Methods (5th ed)*. Oxford University Press, 260pp.

Lawrimore, J. H., and T. C. Peterson, 2000: Pan evaporation trends in dry and humid regions of the United States. *J. Hydrometeorol.*, **1**, 543-546, https://doi.org/10.1175/1525-7541(2000)001<0543>2.0.CO;2..

Liu, B. H., M. Xu, M. Henderson, and W. G. Gong, 2004: A spatial analysis of pan evaporation trends in China. *J. Geophys. Res.*, **109**, D15102, https://doi.org/10.1029/2004JD004511.

Liu, S.-D., F.-M. Liang, S.-S. Liu, and G.-J. Xin, 2003: *Chaos and Fractal in Science*. Peking University Press, 159pp.

Lorenz, E. N., 1963: Deterministic nonperiodic flow. *J. Atmos. Sci.*, **20**, 130-141, https://doi.org/10.1175/1520-0469(1963)020<0130>2.0.CO;2.

Loria A, Panteley E, Nijmeijer H, 1998: *Control of the chaotic duffing equation with uncertainty in all parameters*. IEEE Transactions on Circuits and Systems-I: Fundamental Theory and Applications 45, 1252-1255, https://doi.org/10.1109/81.736558.

Odenbaugh, J., 2005: Idealized, inaccurate, and successful: A pragmatic approach to evaluating models in theoretical ecology. *Biol. Philos.*, **20**, 231-255, https://doi.org/10.1007/s10539-004-0478-6.

Paltridge, G. W., 1974: Global cloud cover and earth surface temperature. *J. Atmos. Sci.*, **31**, 1571-1576, https://doi.org/10.1175/1520-0469(1974)031<1571>2.0.CO;2.

Peterson, T. C., V. S. Golubev, and P. Ya. Groisman, 1995: Evaporation losing its strength. *Nature.*, **377**, 687-688, https://doi.org/10.1038/377687b0.

Priestley, C. H., and R. J. Taylor, 1972: On the assessment of surface heat flux and evaporation using large-scale parameters. *Mon. Wea. Rev.*, **100**, 81-92, https://doi.org/10.1175/1520-0493(1972)100<0081>2.3.CO;2.

Pykh, Y. A., 2001: Lyapunov functions for lotka-volterra systems: An overview and problems. *Proc. 5th IFAC Symposium on Nonlinear Control Systems, **34**. 2001*, St Petersburg, Russia, 1549-1554, https://doi.org/10.1016/S1474-6670(17)35410-1.

Pykh, Y. A., 2003: Energy lyapunov function for generalized replicator equations. *Proc. Int. Conf. "Physics and Control"*, Petersburg, Russia, 270-275, https://doi.org/10.1109/PHYCON.2003.1236830.

Qian, Y., D. P. Kaiser, R. Leung, and M. Xu, 2006: More frequent cloud-free sky and surface solar radiation in China from 1955 to 2000. *Geophys. Res. Lett.*, **33**, L01812, https://doi.org/10.1029/2005GL024586.

Qian, Y. P., and Y. Y. Huang, 1990: Analysis of affecting factors for soil and ground surface temperatures. *Sci. Metrol. Sin.*, **10**, 237-247.

Qiu, X. F., C. M. Liu, and Y. Zeng, 2003: Changes of pan evaporation in the recent 40 years over the Yellow River basin. *J. Nat. Resour.*, **18**, 437-442, https://doi.org/10.1080/02508060408691814.

R Development Core Team, 2010: A language and environment for statistical computing. R Foundation for Statistical Computing, Vienna, Austria, 1956 pp, http://www.R-project.org/.

Ren, G. Y., and J. Guo, 2006: Change in pan evaporation and influential factors over China: 1956-2000. *J. Nat. Resour.*, **21**, 31-44, https://doi.org/10.11849/zrzyxb.2006.01.005.

Rial, J. A., 2004: Abrupt climate change: Chaos and order at orbital and millennial scales. *Global Planet. Change.*, **41**, 95-109, https://doi.org/10.1016/j.gloplacha.2003.10.004.

Roderick, M. L., and G. D. Farquhar, 2002: The cause of decreased pan evaporation over the past 50 years. *Science.*, **298**, 1410-1411, https://doi.org/10.1126/science.1075390-a.

Roderick, M. L., and G. D. Farquhar, 2004: Changes in Australian pan evaporation from 1970 to 2002. *Int. J. Climatol.*, **24**, 1077-1090, https://doi.org/10.1002/joc.1061.

Rossow, W.B., and E. Duenas, 2004: The International Satellite Cloud Climatology Project (ISCCP) web site: An online resource for research. *Bull. Amer. Meteorol. Soc.*, **85**, 167-172, doi:10.1175/BAMS-85-2-167.

Senior, C. A., and J. F. B. Mitchell, 1993: Carbon dioxide and climate: The impact of cloud parameterization. *J. Climate.*, **6**, 393-418, https://doi.org/10.1175/1520-0442(1993)006<0393:CDACTI>2.0.CO;2.

Sykes, L. R., B. E. Shaw, and C. H. Scholz, 1999: Rethinking earthquake prediction. *Pure Appl. Geophys.*, **155**, 207-232, https://doi.org/10.1007/s000240050263.

Wang, Y. J., T. Jiang, C. Y. Xu, and Y. F. Shi, 2005: Trends of evapotranspiration in the Yangtze River basin in 1961-2000. *Adv. Clim. Change Res.*, **1**, 99-105.

Young, A. R., and J. Sabburg, 2006: Cloud effects on evaporation at sub-tropical site, Australian *Institute of Physics 17th National Congress 2006.*, Brisbane, Australia, http://www.aip.org.au/Congress2006/552.pdf.